\documentstyle[aps]{revtex}
\tightenlines
\def\RE {I\kern-6pt R    }
\def\Z  {Z\kern-13pt Z   }
\def\be {\begin{equation}}
\def\ee {\end{equation}  }
\def\beq{\begin{eqnarray}}
\def\eeq{\end{eqnarray}  }

\def\bi {\begin{itemize} }

\def\ei {\end{itemize}   }

\def\gtwid{\mathrel{\raise.3ex\hbox{$>$\kern-.75em\lower1ex\hbox{$\sim$}}}}
\def\ltwid{\mathrel{\raise.3ex\hbox{$<$\kern-.75em\lower1ex\hbox{$\sim$}}}}

\input epsf

\begin{document}


\title{Critical phenomena at the threshold of black hole formation for collisionless matter in spherical symmetry}

\author{
Ignacio (I\~naki) Olabarrieta$^1$\footnote{{\tt inaki@physics.ubc.ca}}
\ and
Matthew W.~Choptuik$^{1,2,3}$\footnote{{\tt choptuik@physics.ubc.ca}}
}
\address{
${}^{1}$Department of Physics and Astronomy, University of British Columbia,\\
Vancouver BC, V6T 1Z1 Canada\\
${}^{2}$CIAR Cosmology and Gravity Program\\
${}^{3}$Center for Relativity, University of Texas at Austin, TX 78712-1081 USA}
\maketitle

\begin{abstract}

We perform a numerical study of the critical regime at the threshold 
of black hole formation in the spherically symmetric, general relativistic
collapse of collisionless matter. The coupled Einstein-Vlasov equations 
are solved using a particle-mesh method in which the evolution of 
the phase-space distribution function is approximated by a set of 
particles (or, more precisely, infinitesimally thin shells) moving 
along geodesics of the spacetime.  Individual particles may have non-zero
angular momenta, but spherical symmetry dictates that the {\em total}
angular momentum of the matter distribution vanish.  In accord with 
previous work by Rein {\em et al}~\cite{Rein}, our results indicate 
that the critical behavior in this model is Type I; that is, the 
smallest black hole in each parametrized family has a {\em finite} mass.
We present evidence that the critical solutions are characterized by 
unstable, {\em static} spacetimes, with non-trivial distributions of 
radial momenta for the particles.  As expected for Type I solutions, 
we also find power-law scaling relations for the lifetimes of 
near-critical configurations as a function of parameter-space distance 
from criticality. 

\end{abstract}



\section{Introduction}
\label{sec:introduction}
Critical phenomena at the threshold of black hole formation were 
originally discovered  in studies of the spherically symmetric,
general relativistic collapse of a minimally coupled scalar 
field~\cite{Choptuik}.
Similar behavior has now been found in many different scenarios, including the 
collapse of gravitational waves, perfect fluids, Yang-Mills fields and 
scalar fields in anti-de Sitter spacetime (for a review see \cite{Gundlach}). 
Relatively little work has been done on the critical collapse of 
collisionless matter.  To date, the only detailed study of the 
black-hole threshold in the Einstein-Vlasov model is due to 
Rein, Rendall and Schaeffer~\cite{Rein}.  These authors found evidence that 
for spherically symmetric collapse with non-zero angular momenta 
distributions, the threshold black hole mass is {\em finite} (Type I 
behavior).  In this paper we summarize the results of \cite{msc_thesis} 
which corroborate and extend the previous work of Rein~{\em et al.}

The paper is organized as follows:
In section \ref{sec:equations}, 
we outline the specific form of the Einstein-Vlasov equations we 
have solved, and make some contact with the particle-mesh (PM) method
which is subsequently used to numerically solve these equations. Here we
follow the approach of Shapiro and Teukolsky~\cite{Shapiro:1985a}, 
which has been
successfully used to model the dynamics of spherically-symmetric, 
relativistic clusters of stars~\cite{Shapiro:1985b},\cite{Shapiro:1985c},\cite{Shapiro:1986}.
Section \ref{sec:numerical} describes our numerical techniques
{\em per se}, while section \ref{sec:results} contains our main results, 
including evidence that the critical solutions in this model 
are characterized by {\em static} geometries and satisfy the 
type of scaling expected of Type I solutions.
Finally, some brief concluding remarks are made in 
section~\ref{sec:discussion}.

We use geometric units, $G=c=1$, throughout the paper.  Abstract spacetime
indices are generally denoted by $a$ and $b$, while 
$\mu, \nu$ and $k, l$ are used for spacetime and spatial component indices,
respectively.
Finally, subscript $i$'s label specific particles, while subscript $j$'s
are generally used for finite-difference indexing.

\section{Formalism and equations of motion}
\label{sec:equations}
The dynamical state of collisionless matter
 can be described by a distribution function, $f(x^a,p_a)$:
\begin{equation}
\label{distri}
f(x^a,p_a)=dN/dV_p
\end{equation}
where $N$ is the particle number and $V_p$ is the phase-space volume.
In the current case, the volume in phase space
is conserved during the evolution of the system (Liouville's theorem).
This implies that the distribution function is also a conserved quantity:
\begin{equation}
\label{vlasov}
\frac{df(t,x^k,p_k)}{dt}=0
\end{equation}
This is the collisionless Boltzmann, or Vlasov, equation.
This equation, coupled to Einstein's equations, $G_{ab}=8\pi T_{ab}$,
all restricted to spherical symmetry, i.e.:
\begin{equation}
f(t,x^k,p_k)=f(t,R \, x^k,R\, p_k) \quad {\mathrm{with}}\quad R \in SO(3) \quad k=1,2,3
\end{equation}
form the system we wish to solve numerically.

\subsection{Maximal-areal coordinate system.}
As with any problem in ``3+1'' (``space + time'') numerical relativity, 
we want to specify initial data on a spatial 
hypersurface and then evolve these data in time. 
To do this, we need to split the Einstein equations  
into a set of constraint equations (equations that must be 
satisfied at each instant of time) and dynamical or evolution
equations (equations that tell us how to evolve the geometric 
quantities in time). 
We carry out this splitting using the 3+1 formalism due to Arnowitt,
Deser and Misner (ADM) (for reviews of this formalism, see \cite{York} and \cite{MTW}). 

We restrict attention to spherical symmetry and adopt coordinates 
$(t,r,\theta,\varphi)$ with the usual spherical-polar topology. 
We are left with the freedom to choose our radial and time coordinates,
and have chosen maximal-areal coordinates. As the name suggests, in 
this system the radial coordinate is areal, so that
the proper area of 2-spheres with radius $r$ is $4\pi r^2$.
The time coordinate is fixed by
demanding that the $t={\rm constant}$, 3-slices be {\em maximal}, i.e. 
that the trace of the extrinsic curvature, $K\left(t,r\right) 
\equiv K^l{}_l(t,r)$, identically vanish on each slice.
This leads to a {\em slicing condition} on the {\em lapse function},
$\alpha\left(t,r\right)$, which must be satisfied at each instant
of time. 

With these choices, the spacetime metric takes the specific form:
\begin{equation}
\label{metric}
ds^2=\left(-\alpha(t,r)^2+a(t,r)^2\beta(t,r)^2\right)dt^2+
     2a^2\beta\, dtdr+a^2\,dr^2+r^2\left(d\theta^2 + \sin^2\theta d\varphi^2\right)
\end{equation}
where $\beta(t,r)$ is the radial component of the {\em shift vector}, 
$\beta^k = (\beta,0,0)$. A sufficient set of equations for determining 
the geometric quantities, $a(t,r)$,
$K^\theta{}_\theta(t,r)$, $\alpha(t,r)$ and $\beta(t,r)$
is then:
\ \\
{\it Hamiltonian Constraint:} 
\begin{equation}
\label{geo_eqn1}
\frac{a^\prime}{a}=\frac{3}{2}a^2r{K^\theta{}_\theta}^2+4\pi r a^2\rho+\frac{1}{2r}
\left(1-a^2\right)
\end{equation}
{\it Momentum Constraint:} 
\begin{equation}
{K^\theta{}_\theta}^\prime=-\frac{3}{r}K^\theta{}_\theta-4\pi j_r
\end{equation}
{\it Slicing Condition:}
\begin{equation}
\alpha^{\prime \prime}=\alpha^\prime\left( \frac{a^\prime}{a}- \frac{2}{r}\right)+
     \frac{2 \alpha}{r^2}\left( 2r \frac{a^\prime}{a}+a^2-1 \right)+4 \pi a^2 \alpha
\left(S-3\rho \right)
\end{equation}
{\it Areal Coordinate Condition:}
\begin{eqnarray}
\label{geo_eqn2}
\beta&=&\alpha r K^\theta{}_\theta
\end{eqnarray}
Here, ${}'$ denotes differentiation with respect to $r$, and 
the last formula is derived from $\partial_t K = 0$, using $K=0$.
In addition, $\rho(t,r)$, $j_r(t,r)$ and $S(t,r)$, which are 
discussed in detail
in the next section, are the local 
energy density, the local current density and the trace of the spatial 
part of the stress-energy density, respectively.
We note that we have chosen to implement a {\em fully constrained} 
evolution, which in this case means
that we use the constraint equations, rather than
evolution equations, to update $a$ and $K^\theta{}_\theta$.

During our simulations, we also compute the {\em mass aspect
function}, $M(t,r)$: 
\begin{equation}
\label{mass}
M(t,r)=\frac{r}{2}\left(1+\frac{\beta^2}{\alpha^2}-\frac{1}{a^2}\right)=
\frac{r}{2}\left(1+r^2K^\theta{}_\theta{}^2-\frac{1}{a^2}\right),
\end{equation}
which, among other useful diagnostic purposes, allows us
detect the formation of apparent horizons.
Specifically, when 
$2M(t,r)/r=1-1/a^2+\beta^2/\alpha^2\,$, becomes equal to 1,
a marginally trapped surface has been formed.
We can see this by computing the expansion of the outgoing null geodesics
(see for instance \cite{Choptuik:phd}), which in these coordinates can be written as:
\begin{equation}
1-a(t,r)\,r\,K^\theta{}_\theta(t,r)\,=\,1-\frac{a(t,r)\,\beta(t,r)}{\alpha(t,r)}.
\end{equation}
Therefore, if the outgoing expansion is zero, $1/a^2 = \beta^2/\alpha^2$, and 
$2M(t,r)/r = 1$.
\subsection{Stress-energy tensor}
In this section we explain how we calculate the stress-energy quantities 
$\rho(t,r)$, $j_r(t,r)$
and $S(t,r)$ that appear in equations (\ref{geo_eqn1}-\ref{geo_eqn2}). 
Adopting a Monte Carlo approach, we approximate the
distribution function (\ref{distri}) by a set of $N$ ``spherical particles'',
which actually represent infinitesimally thin spherical shells of matter.
Since these particles only interact with each other gravitationally, we have
\begin{equation}
T^{\mu \nu}=\sum _{i=1}^N T^{\mu \nu}_i
\end{equation}
where $T^{\mu \nu}_i$ is the stress energy tensor for a single particle.
For a point particle we have
\begin{equation}
\label{tmunu}
T^{\mu \nu}_i =\frac{p^\mu_i\, p^\nu_i}{m_i}
\delta(\vec{r}-\vec{r}_i(t)),
\end{equation}
where, $p^\mu _i$ are the components of the 4-momentum of the $i$-th
particle, $m_i$ is its rest mass, $\vec r_i(t)$ is its radial position at 
time $t$, and $\delta$ is the usual Dirac $\delta$-function.
In maximal-areal coordinates, the single-particle contributions to the
quantities $\rho$, $S$ and $j_r$ 
then take the form:
\begin{eqnarray}
\left[\rho\right]_i&=&\alpha^2 \left[T^{tt}\right]_i\\
\left[S\right]_i&=&\frac{1}{a^2} \left[T_{rr}\right]_i +
\frac{1}{r^2}\left[T_{\theta\theta}\right]_i + \frac{1}{r^2 {\mathrm sin}^2{\theta}}\left[T_{\varphi\varphi}\right]_i\\
\left[j_r\right] _i&=&\alpha \left[T^t{}_r\right] _i .
\end{eqnarray}
We now relax the point particle approximation and assume that each particle
is a spherically symmetric shell of mass, uniformly distributed over a
region $\Delta r$ in radius (subsequently,
$\Delta r$ will be identified with  the mesh spacing, $h$, 
used in the finite-difference solution of the geometrical equations). 
Each shell of matter is to be 
interpreted as an average over
an ensemble of shells, each centered  at $r=r_i$, and with angular momentum
vectors which point in all possible directions. 
Thus, for any shell, the net angular momentum is zero, 
$\vec l=0$, but $|{\vec l}|^2 \equiv l^2 \neq 0$.
The proper volume occupied by each particle is then given by:
\begin{equation}
V_i = \frac{p^t}{m_i}\,\Delta r\int \sqrt{-g}\,\, d\varphi\, d\theta
    =4 \pi\, \Delta r\, \alpha\, a\, \frac{r^2_i \,\, p^t_i}{m_i},
\end{equation}
and we can approximate the delta function that appears in
(\ref{tmunu}) by $1/V_i$.
This yields:
\begin{eqnarray}
\rho_i&=&\frac{1}{4\pi\Delta r a} \frac{\left[\bar p^t\right]_i}{r^2_i}\\
\label{sr_sa}
S_i=\left[S^r{}_r \right]_i+\left[S^a{}_a \right]_i&=&\frac{1}{4\pi\Delta r a^3 }
    \frac{{\left[p_r\right]_i}^2}{\left[\bar p^t\right]_ir^2_i}+
    \frac{1}{4\pi\Delta r a} \frac{\left[l\right]^2_i}{r^4_i \left[\bar p^t\right]_i}\\
\left[ j_r \right] _i&=&\frac{1}{4\pi\Delta r a} \frac{\left[p_r\right]_i}{r^2_i}.
\end{eqnarray}
Here $\left[\bar p^t\right]_i$ is defined by $\left[\bar p^t\right]_i \equiv \alpha \left[p^t\right]_i$, and
${\left[l\right]_i}^2 \equiv {\left[p_\theta \right]_i}^2+{\left[p_\varphi\right]_i}^2 / \sin^2{\theta_i}$
is the square of the magnitude of the angular momentum of the $i$-th particle.
The geometric quantities $\alpha$ and $a$ are evaluated at $r=r_i$ as 
described in~Sec.~\ref{interpolation}.
Note that we have also defined
\begin{equation}
S^a{}_a \equiv S^\theta{}_\theta + S^\varphi{}_\varphi
\end{equation}
i.e., for $S^a{}_a$ the index $a$ is summed over the {\em angular} coordinates.
We then introduce quantities which do not explicitly
depend on the geometrical quantities:
$\left[ \bar \rho \right]_i \equiv a \left[\rho \right]_i$,
$\left[\bar S^r{}_r\right]_i \equiv a^3 \left[S^r{}_r \right]_i$, $\left[ \bar S^a{}_a\right]_i \equiv a
\left[S^a{}_a \right]_i$ and $\left[\bar j_r\right]_i \equiv  a \left[j_r\right]_i $.
In our numerical implementation of the equations of motion, these 
definitions provide a clean separation of the particle updates and 
the updates of the geometry variables.

We interpolate the one-particle quantities to the continuum
and sum over all the particles to find the total values:
\begin{equation}
\label{inter}
\bar f(r)=\sum_{i=1}^{N} \bar f_i \, W(r-r_i),
\end{equation}
where $\bar f_i$ is any of the single-particle barred quantities defined above,
$\bar f$ is the corresponding continuum quantity, and $W(r-r_i)$ is an
interpolation function defined in detail in Sec.~\ref{interpolation} 
(see equation~(\ref{kernel})).
Having defined (\ref{inter}) we can now write equations (\ref{geo_eqn1}-\ref{geo_eqn2}) as
\begin{eqnarray}
\label{field1}
\frac{a'}{a}&=&\frac{1-a^2}{2r}+\frac{3}{2}ra^2 {K^\theta{}_\theta}^2 + 4\pi
a r \bar \rho\\
 \label{field2}
{K^\theta{}_\theta}'&=&-\frac{3}{r}K^\theta{}_\theta-4\pi \frac{\bar j_r}{a}\\
\alpha''&=&\alpha'\left(\frac{a'}{a}-\frac{2}{r}\right)+\frac{2\alpha}{r^2}
\left(a^2-1+2r\frac{a'}{a}\right)+4\pi
a \alpha \left(\frac{\bar S^r{}_r}{a^2}+\bar S^a{}_a-3 \bar \rho \right)\\
\label{field4}
\beta&=&\alpha r K^\theta _\theta
\end{eqnarray}
\subsection{Evolution equations}
Because there are no explicit interactions between the particles, 
their equations of motion are just the spacetime geodesic equations
(the characteristics of the Vlasov  equation). These can be derived 
from the formula for parallel transport of a particle's four-momentum
along its world line:
\begin{equation}
\label{geodesic}
p^a \nabla _a p^b = 0.
\end{equation}
It proves useful to recast these equations in terms of the 
quantities, $p_r$, $p^t$ and 
${l}^2 \equiv {p_\theta }^2+{p_\varphi}^2 / \sin^2{\theta}$.
We can express $p^r$ in terms of these variables as:  
\begin{equation}
\label{p^r}
p^r(t,r)=\frac{p_r(t,r)}{a^2(t,r)}-\beta(t,r)\,p^t(t,r)
\end{equation}
To compute total derivatives with respect to coordinate time
we use 
\begin{equation}
\frac{d}{d t}=\frac{\partial }{\partial t}+
                 \frac{d r}{d \tau}
                 \frac{d \tau}{d t}\frac{\partial }{\partial r}
             =\frac{\partial }{\partial t}+
                 \frac{p^r}{p^t}\frac{\partial }{\partial r},
\end{equation}
where here, and in the remainder of this section, $\tau$ is the 
particle's proper time.
Applying this operator to equation (\ref{p^r}) we get
\begin{equation}
\label{dp^rdt2}
\frac{dp^r}{dt}=\frac{1}{a^2}\frac{dp_r}{dt}
                 -2\frac{p^r}{a^3}\left(\frac{\partial a}{\partial t}
                 +\frac{p^r}{p^t}\frac{\partial a}{\partial r}\right)
                 -\beta \frac{dp^t}{dt} - p^t\left(
                 \frac{\partial \beta}{\partial t} + \frac{p^r}{p^t}
                 \frac{\partial \beta}{\partial r}\right).
\end{equation}
Substituting equations (\ref{p^r}) and (\ref{dp^rdt2}) into equation
(\ref{geodesic}) we obtain:
\begin{equation}
\label{dprdt3}
\frac{d p_r}{d t} =-\alpha \frac{\partial\alpha}{\partial r} p^t+
\frac{\partial \beta}{\partial r} p_r +
\frac{1}{a^3}\frac{\partial a}{\partial r}\frac{{p_r}^2}{p^t}+
\frac{l^2}{p^t r^3},
\end{equation}
which is the evolution equation for $p_r$. To derive the evolution equation
for $r$, we use the definition of $p^r$ ($p^r \equiv dr/d\tau$), 
which after some manipulation yields:
\begin{equation}
\label{drdt3}
\frac{dr}{dt}=\frac{p_r}{a^2 p^t}-\beta
\end{equation}
The time component of the 4-momentum, $p^t$, is calculated using the 
normalization condition $p^\mu p_\mu=-m^2$:
\begin{equation}
\label{alpt3}
\alpha p^t = \sqrt{m^2+\frac{{p_r}^2}{a^2}+\frac{l^2}{r^2}}
\end{equation}
It is also convenient, as previously mentioned, to use $\bar p^t=\alpha p^t$
rather than $p^t$ itself.  Using this definition in 
equations~(\ref{dprdt3}), (\ref{drdt3}) and~(\ref{alpt3}) yields the 
final form of the particle equations of motion:
\begin{eqnarray}
\label{evol1}
\frac{d p_r}{d t} &=&-\frac{\partial\alpha}{\partial r} \bar p^t+
\frac{\partial \beta}{\partial r} p_r +
\frac{\alpha}{a^3}\frac{\partial a}{\partial r}\frac{{p_r}^2}{\bar p^t}+
\frac{l^2 \alpha}{\bar p^t r^3}\\
\label{evol2}
\frac{d r}{d t}&=&\frac{\alpha p_r}{a^2 \bar p^t}-\beta\\
\label{evol3}
\bar p^t& =& \sqrt{m^2+\frac{{p_r}^2}{a^2}+\frac{l^2}{r^2}}.
\end{eqnarray}

\section{Numerical approach}
\label{sec:numerical}

As discussed in the previous section, we have adopted a Monte Carlo,
particle-based strategy to the solution of the Vlasov equation. 
In this approach we generate an $N$-particle sample of some
specified initial distribution function $f(0,x^k,p_k)$, and then
use dynamical evolution of the $N$ particles to approximate the full 
dynamics of $f(t,x^k,p_k)$.  The continuum limit is recovered in the limit 
$N\to\infty$ and, in the absence of any sophisticated ``importance sampling'' techniques, we expect the level of statistical error in 
our particle calculations to be of the order of $1/\sqrt{N}$.
We couple the particles to the gravitational field by introducing 
a finite-difference mesh on which we approximately solve the geometric 
equations, and by introducing transfer operators which allow us 
to produce mesh-based representations of particle quantities and 
{\em vice versa}.  ``Particle-Mesh'', or PM, methods such as ours
are commonly used in the solution of Boltzmann equations, particularly 
those involving long-range interactions, and the reader is referred 
to~\cite{Hockney} for a detailed review of such techniques.

Here we simply note that a PM method is generically characterized by 
the splitting of each discrete time step, $t^n \to t^{n+1}$  into two stages:
1) the solution of the field equations on a finite-difference mesh, and 
2) the updating of particle positions via discrete versions of their 
equations of motion.  In our case, and as described in the 
previous section, at each time step the stress-energy quantities are 
calculated by considering each particle 
to be ``smoothed'' over a finite volume.

\subsection{The field equations}
We first explain how the equations (\ref{field1}-\ref{field4}) for the geometry
are solved numerically, assuming that we know the quantities
$\bar \rho$, $\bar j_r{}$, $\bar S^r{}_r{}$, $\bar S^a{}_a$ (our 
computation of the stress-energy quantities is described in 
Section~\ref{interpolation}).
The first two equations, equations
(\ref{field1}-\ref{field2}), are
integrated from the origin, $r=0$, using the {\tt lsoda} \cite{lsoda}
integrator. The boundary conditions are given by the spherical symmetry of the
spacetime, and by the demand that the spacetime be locally flat at $r=0$.
They are $a(t,0) = 1$, and $K^\theta{}_\theta(t,0)=0$.

We compute the values of the functions $a_j$ and
$K^\theta{}_\theta{}_j$ on a uniform grid of $N_r$ points, 
$r_j \equiv (j-1) h, \,\, j = 1, \, \cdots\, N_r$, where $h \equiv \Delta r = 
r_{\rm max} / (N_r - 1)$, and $r = r_{\rm max}$ is the outer edge of the 
computational domain.

In order to compute the values at $r=r_{j+1}$, we supply to {\tt lsoda} the values of the
functions at $r=r_j$ and the derivatives computed using equations (\ref{field1}-\ref{field2})
at $r=r_{j+1/2}$, using the average of $\bar \rho$ and $\bar j_r$ at $r_j$ and 
$r_{j+1}$.
\begin{eqnarray}
\label{rhoaver}
\left[ \bar \rho \right] _{j+1/2} &=& \frac{1}{2} \left( \left[\bar \rho\right]_j + \left[\bar \rho\right]_{j+1} \right)\\
\label{javer}
\left[\bar j_r\right]_{j+1/2} &=& \frac{1}{2} \left( \left[\bar j_r\right]_j + 
\left[\bar j_r\right]_{j+1} \right).
\end{eqnarray}
Once we have calculated $a$, we can solve the slicing equation:
\begin{equation}
\label{sli}
\alpha''=\alpha'\left(\frac{a'}{a}-\frac{2}{r}\right)+\frac{2\alpha}{r^2}\left(a^2-1+2r\frac{a'}{a}\right)+
4\pi a \alpha \left(\bar S^r{}_r+\bar S^a_a-3 \bar \rho\right),
\end{equation}
with the boundary conditions:
\begin{eqnarray}
\alpha^\prime(t,0)&=&0,\\
\label{alphabcrmax}
\alpha(t,\infty)&=&1.
\end{eqnarray}
Here the first condition follows from the demand that the slicing be 
regular at $r=0$,
and the second one follows from asymptotic
flatness, plus the demand that $t$ measure proper time at infinity.
We solve~(\ref{sli}) using a second-order finite-difference approximation on the 
finite-difference mesh:
\begin{eqnarray}
\nonumber
\frac{\alpha_{j+1}-2 \alpha_j+\alpha_{j-1}}{h^2}=&&\frac{\alpha_{j+1}-\alpha_{j-1}}{2 h}
\left(\frac{a_{j+1}-a_{j-1}}{2 h a_j} - \frac{2}{r_j}\right) +\\ \nonumber
 && \frac{2 \alpha_j}{{r_j}^2}
\left( {a_j}^2 - 1 +2 r_j \frac{a_{j+1}-a_{j-1}}{2 h a_j}\right)+\\ 
&&4 \pi a_j \alpha_j \left( \frac{\left[\bar S^r{}_r\right]_j}{{a_j}^2} +\left[\bar S^a_a\right]_j -
3\left[\bar \rho \right]_j\right)
\end{eqnarray}
Rearranging this equation gives us:
\begin{equation}
\label{slieq}
\left( \frac{1}{h^2} + \frac{f_j}{2h} \right) \alpha_{j-1}-
\left( \frac{2}{h^2} + g_j \right) \alpha_j+
\left( \frac{1}{h^2} - \frac{f_j}{2h}\right) \alpha_{j-1}=0
\end{equation}
where:
\begin{eqnarray}
f_j&=& \frac{a_{j+1}-a_{j-1}}{2ha_j} - \frac{2}{r_j}\\
g_j&=& \frac{2}{{r_j}^2}\left( {a_j}^2-1+2 r_j \frac{a_{j+1}-a_{j-1}}{2 h a_j}\right)+
4 \pi a_j \left( \frac{\left[\bar S^r{}_r\right]_j}{{a_j}^2} +\left[\bar S^a_a\right]_j -
3\left[\bar\rho\right]_j\right)
\end{eqnarray}
In addition to~(\ref{slieq}) we have the boundary equation at $r=0$:
\begin{equation}
\label{slieqi}
\left(-3 + \frac{1/h^2 + f_2/(2 h)}{1/h^2 - f_2/(2 h)} \right) \alpha_1 +
\left(4 + \frac{-2/h^2-g_2}{1/h^2-f_2/(2h)} \right) \alpha_2 = 0,
\end{equation}
which can be derived from the $O(h^2)$ forward finite-difference approximation 
to $\alpha^\prime=0$ at $r=0$
\begin{equation}
\frac{-3 \alpha_1 + 4 \alpha_2 - \alpha_3}{2h}=0\ ,
\end{equation}
and equation (\ref{slieq}) with $j=2$. We have also the boundary condition at
$r=r_{\mathrm max}$:
\begin{equation}
\label{l_rmax}
\alpha_{N_r} = \sqrt{1 - \left(\frac{2 M_{N_r}}{r_{N_r}}\right)}
\end{equation}
where $M$ is the mass aspect function defined by equation~(\ref{mass}).
This approximation follows from the known representation of the 
asymptotically-flat Schwarzschild solution in maximal-areal coordinates.
Equations~(\ref{slieq}), (\ref{slieqi}) and (\ref{l_rmax}) constitute a 
linear tridiagonal system 
that can be solved using a
tridiagonal solver (we have used the LAPACK \cite{LAPACK} routine {\tt dgtsv}).
\\
\subsection{The evolution equations}
\label{evolution}
To evolve the particles' positions and momenta we integrate the geodesic equations
(\ref{evol1}-\ref{evol2}). The values of the coefficients in these equations
(basically products and quotients of $a$, $\alpha$, $\beta$, $a^\prime$, $\alpha^\prime$
 and $\beta^\prime$)
must be calculated at the particle positions, $r_i$, using the values 
obtained at the mesh points, $r_j$.
The mesh values are interpolated to the particle positions using the same operator kernel
used to produce mesh values from particle quantities (this procedure is explained
in the next section). The geodesic equations are also integrated 
using the {\tt lsoda}
routine.
At discrete time $t=t^n$, given a particle's position, $r^n$, and radial
momentum, $p_r^n$, we calculate the new position,
$r^{n+1}$, and momentum, $p_r^{n+1}$, at $t=t^{n+1} = t^n + \Delta t$ by
supplying to {\tt lsoda} the values
of the metric functions and their spatial derivatives evaluated at $t=t^n$. 
Because we use the $t=t^n$ values of the geometric quantities in the 
particle updates, rather than, for example, values at $t=t^{n+1/2}$,
we expect our solution of the particle equations to have accuracy 
$O(\Delta t)$.
We also note that in our numerical implementation, we chose a value of 
$\Delta t$ proportional to $h$, i.e. $\Delta t = \lambda\, h$, 
where usually $\lambda = 1.0$.

We need to take special care if a particle leaves the computational 
domain
($r_i>r_{\mathrm max}$) or if it reaches the origin. In the first case we simply 
remove the particle from the integration scheme.
When a particle reaches the origin, which operationally is signaled
by $r_i<0$, we ``reflect''
the particle by setting:
\begin{eqnarray}
r_i  &\to& -r_i\,,\\
\left[p_r \right]_i &\to& -\left[ p_r \right]_i\,,\\
l_i &\to& l_i
\end{eqnarray}
\subsection{Interpolation and restriction}
\label{interpolation}
In this section we explain how we calculate the stress-energy quantities 
on the finite-difference mesh
from a given set of particles (restriction), as well as how 
we interpolate the geometric quantities from the finite-difference mesh 
to the particles' positions.

The values of the stress-energy quantities $\bar \rho$, $ \bar j_r$, $\bar S$ 
are calculated on the mesh using equation (\ref{inter}):
\[
\bar f(r_j) = \sum_{i=1}^{N} \bar f_i \  W(r_j-r_i)
\]
where $\bar f_i$ are the single particle quantities.
In our implementation, we use the specific kernel:
\begin{equation}
\label{kernel}
W(r_j-r_i)= \left\{ \begin{array}
{r@{\quad:\quad}l}
1-|r_j-r_i|/h & |r_j-r_i| \le h \\ 0 & {\mathrm {otherwise}} \end{array} \right.
\end{equation}
Similarly, to restrict the geometric quantities
calculated on the mesh to the particles' positions we compute:
\begin{equation}
\label{restr}
F(r_i)=\sum^{N_r}_{j=1} F(r_j) W(r_j-r_i)
\end{equation}
where, again, $r_i$ is the position of the particle, $r_j$ are
the grid points, and $F$ is any of the coefficients which appear in 
the geodesic equations.
These coefficients are generally products and quotients of metric functions 
and their derivatives.
In order to calculate derivatives we use the standard $O(h^2)$ 
finite-difference approximation:
\begin{equation}
\left[F^\prime\right]_j=\left( F_{j+1}-F_{j-1} \right) /(2 h) + O(h^2)
\end{equation}
and then use equation (\ref{restr}) to find an approximate value for
$F^\prime(r_i)$.

\subsection{Initial data}
To initialize the sets of particles which we evolve, we specify the
particle distribution (number of particles per unit of areal coordinate)
and the velocity distribution 
(specifically the number of particles per $p_r$ and $l$).
This corresponds to a {\em separable} distribution function,
$f(r,p_r,l)$:
\begin{equation}
\label{distris}
f(r,p_r,l)=R(r)\,P(p_r)\,L(l)
\end{equation}

Moreover, instead of specifying $P$ as a function of $p_r$ we give 
$P=P(\bar p_r)$ where $\bar p_r=p_r/a$. 
This allows us
to calculate the value of $p^t=\sqrt{m^2+\bar p_r^2 +l^2/r^2}$ (assuming all the particles have
the same rest mass $m$), and therefore 
$[\bar \rho]_i$, $[\bar j_r]_i$ and $[\bar S]_i$ without {\em a priori}
knowledge of the geometry.
We can thus decouple the tasks of specifying initial conditions for the 
particles, and ensuring that the constraints are satisfied at $t=0$.

We use 1-dimensional Monte Carlo techniques applied to each of 
$R(r)$, $P(p_r)$,
$L(l)$ to get a specific set of $N$ particles.
As mentioned above, the statistical error, in theory, should scale 
as $1/\sqrt{N}$.

\section{Results}
\label{sec:results}
All of the calculations discussed in this paper were 
performed with $N_r=257$, $N=10^5$ and $h=0.078125$.
With this choice of parameters we ensured that the truncation error due to the 
finite-differencing of the field equations with mesh spacing $h$ was of the same 
order of magnitude as the statistical error resulting from representation of the 
phase-space distribution with a finite number, $N$, of particles.
We have observed~\cite{msc_thesis} 
that both types of error scale in the expected way:
the truncation error scales as $O(h)$, with the number of particles per 
cell fixed; the statistical error scales as $1/\sqrt{N}$ for fixed $h$, and as
$O(1/h)$ for fixed $N$.
Once the two errors are of comparable magnitude, in order to
further decrease the overall error (truncation plus statistical error)
as $O(h)$, we have to increase the number of particles as $N \approx 1/h^4$. 
This scaling behavior makes it very costly to substantially reduce the 
level of numerical error in the results presented here.

Our critical solutions (solutions sitting at the threshold of black hole
formation) were found by performing
bisection searches using the total rest mass, $M_0$:
\begin{equation}
M_o=\sum_{i=1}^N m_i = N m
\end{equation}
as the tuning parameter.  In particular, if 
$M_o^\star$ is the critical parameter value, then configurations with 
$M_o < M_o^\star$ (subcritical) will eventually disperse, while 
those with $M_o > M_o^\star$ (supercritical) will form black holes.
For any given critical search, we generally determined $M_o^\star$ 
to a relative precision of about $4 \times 10^{-11}$.
Table~\ref{table} provides a summary of the various families we 
have studied.

\begin{figure}[htbp]
\epsfxsize=4.0in
\centerline{\epsfbox[100 150 550 700]{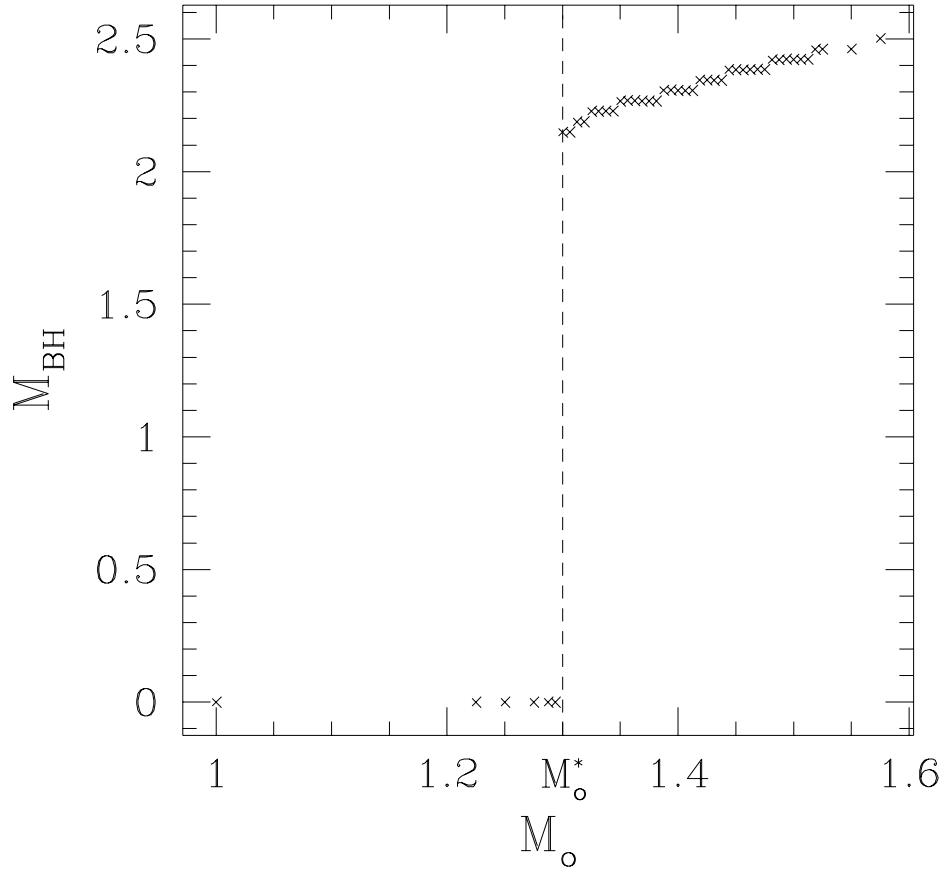}}
\caption[Mass of the black Hole as function of $M_o$.]
{
Black hole mass as function of total rest mass, $M_o$, for the 
``almost time-symmetric'',
gaussian family of initial data described in the text 
(Table~\ref{table}, Family~(a)).
We observe that the smallest black hole has a finite mass (Type I transition)
at a critical parameter $M_o^\star \approx 1.3$ (dashed line).
Computationally, we have tuned $M_o^\star$ to a relative precision of
about $4 \times 10^{-11}$, which is typical of the critical surveys
discussed in this paper.   The discrete jumps in the black hole masses
for $M_o > M_o^\star$ reflect the discrete nature of the finite-difference
grid.  We have made no attempt to ``interpolate'' the location of the black
hole horizon in the finite-difference mesh; hence our mass estimates will
always satisfy $M_{\rm BH} = k\Delta r = kh$, for some integer $k$.
}
\label{MvsMo}
\end{figure}

We first focus on a specific family of initial conditions 
(Table~\ref{table}, Family~(a)) and 
then summarize our observations for the remaining families.
We thus consider an initial distribution defined by:
\begin{eqnarray}
\label{gaussian1}
R(r)&=&r^2 e^{\left(-(r-r_o)^2/\Delta_r^2 \right)}\,\Theta(r)\\
\label{gaussian2}
P(\bar p_r)&=&e^{\left(-(\bar p_r-{\bar p_r}{}_o)^2/
                       {\Delta_{\bar p_r}}^2 \right)}\\
\label{gaussian3}
L(l)&=&e^{\left(-(l-l_o)^2/{\Delta_l}^2 \right)}\,\Theta(l)
\end{eqnarray}
where, again,  $\bar p_r=p_r/a$, and $\Theta$ is the step function. 
We take $r_o=5$, $\Delta_r=1$, ${\bar p_r}{}_o=0$, $\Delta_{\bar p_r}=2$,
$l_o=12$, $\Delta_l=2$, and refer to the data as ``almost time symmetric''
(ATS) since the gaussian for the radial momentum is centered at $p_r=0$.
The critical parameter for this family is $M_o^\star \approx 1.3$ as 
shown in Fig~\ref{MvsMo}.  This figure also shows that the smallest black hole 
formed has {\em finite} mass and that the transition is therefore of 
Type I, in agreement with the observations of Rein {\em et al}~\cite{Rein}.

\begin{figure}[htbp]
\epsfxsize=4.0in
\centerline{\epsfbox[100 150 550 700]{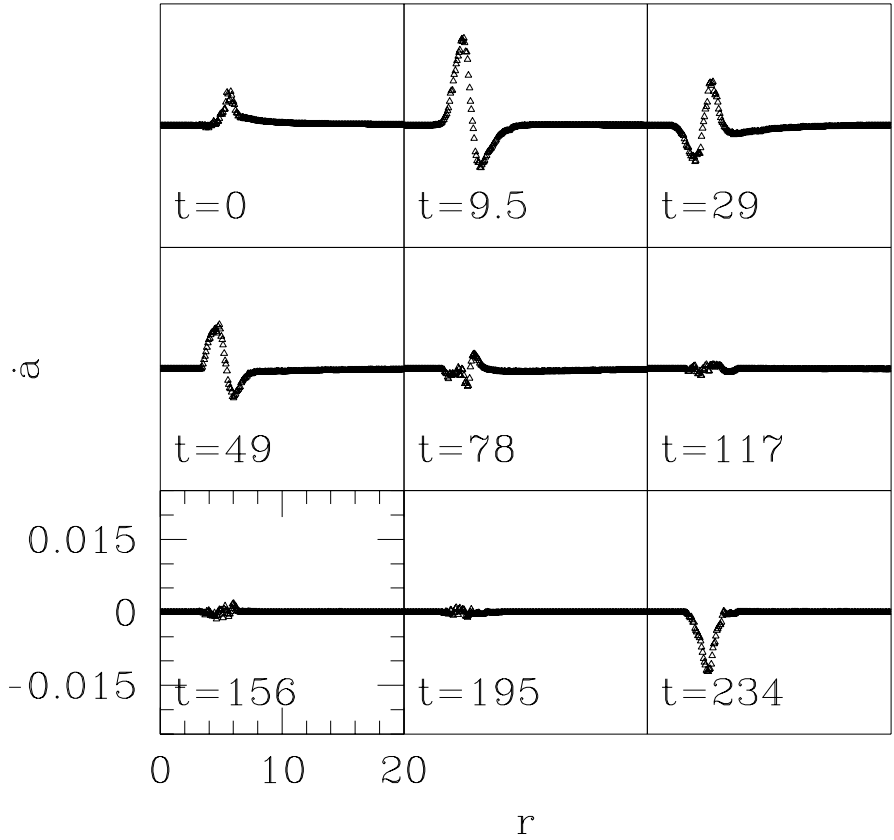}}
\caption[
Evolution of ${\dot a}$ from a marginally subcritical calculation
($M_o < M_o^\star$)
using Family~(a).
Note that at intermediate times $|{\dot a}| \approx 0$.
]
{
Evolution of ${\dot a}$ from a marginally subcritical calculation
($M_o < M_o^\star$)
using Family~(a).
Note that at intermediate times $|{\dot a}| \approx 0$.
}
\label{dadt1}
\end{figure}

In Fig. \ref{dadt1} we show a few snapshots
of the evolution of $\dot a(t,r)\equiv \partial_t a(t,r)$ resulting 
from initial data which is close to criticality but which eventually disperses.
At early times, ${\dot a}(t,r)$ oscillates,
but for $100 \lesssim t \lesssim 200$ it appears to approach 0.
In the last snapshot ($t=234$) we observe that $\dot a(t,r)$ has become 
negative,
corresponding to dispersal of the particles.
We also observe similar behavior for the time derivatives $\dot \alpha$
and $\dot \beta$.

\begin{figure}[htbp]
\epsfxsize=4.0in
\centerline{\epsfbox[100 150 550 700]{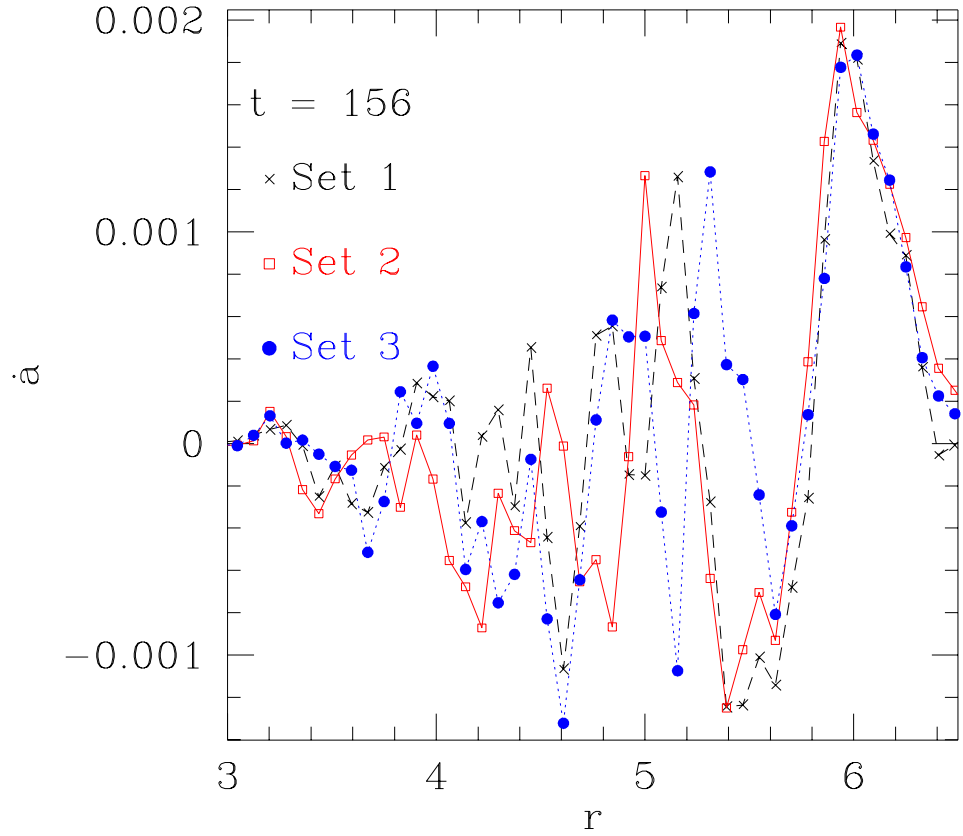}}
\caption[Three ensembles of $\dot a$ at $t=156$.
In the upper figure we can see a detail
of the region where $\dot a$ is not $0$ showing that there is no correlation between the
three ensembles and that the non-zero value is due to statistical noise.]
{
Plot of $\dot a(156,r)$ from three separate Family~(a)
calculations using distinct initial
particle sets ($N=10^5$).  The scatter in the displayed datasets gives
a rough indication of the level of statistical error $\Delta_S({\dot a})$
in the computations. The plot
shows that, at least for $r \lesssim 5.5$,
where $93\%$ of the mass of the putative static cluster is located, there is
little or no correlation between the three sets.
Thus, any non-zero value of ${\dot a}$ in the critical limit may be attributable to
finite-$N$ statistical fluctuations.
}
\label{dadt_156}
\end{figure}

In order to better see  how small $\dot a(t,r)$ becomes, we show in Fig.
\ref{dadt_156} details of $\dot a(t,r)$  at
$t=156$ for three different sets of particles
sampled from the same initial distribution
function. The difference between the solutions obtained with different sets 
give us an estimate of the statistical error, $\Delta_S ({\dot a})$,
in the calculation.
From the figure we can see that for the most part,
$|\Delta_S ({\dot a})| \approx |{\dot a}|$.
This is not the case for $r \gtrsim 5.5$ where the three calculations
all seem to indicate a specific non-zero value for ${\dot a}$; however,
we suspect that the amplitude of this feature may decrease if we tune closer 
to the critical solution, and if we use greater resolution.  More importantly, 
the region $r \gtrsim 5.5$ accounts for 
only about 5-10\% of the mass of the near-critical configuration.

\begin{figure}[htbp]
\epsfxsize=4.0in
\centerline{\epsfbox[100 150 550 700]{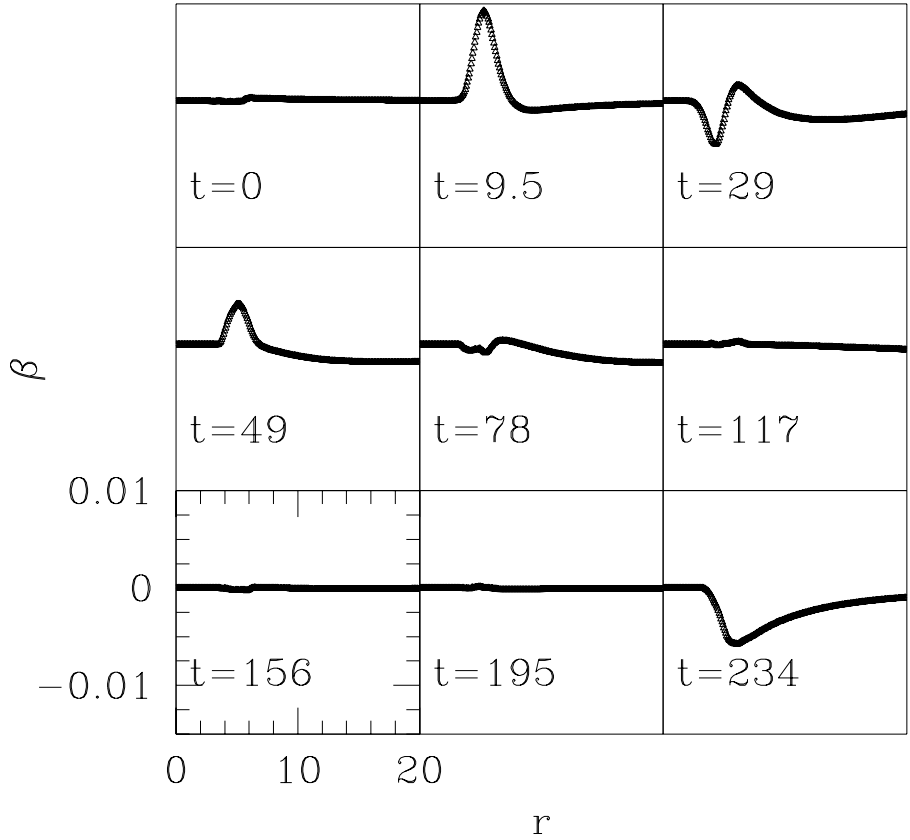}}
\caption[Shift function as a function of time. The shift function becomes
close to zero during the same period of time as does $\dot a$.
$N^a=(\partial / \partial t)^a$ is orthogonal to the
hypersurfaces showing that the geometry is static.]
{
Evolution of $\beta$ from a marginally subcritical calculation
using Family~(a).
The shift function apparently vanishes
during the same period of time as does $\dot a$.
Therefore, during this interval, we have evidence that
$N^a=(\partial / \partial t)^a$ is orthogonal to
the hypersurfaces, and, thus, that the geometry is static.}
\label{b}
\end{figure}

These results are thus {\em consistent} with ${\dot a} \to 0$ in the critical
regime, although more definitive proof would require significantly higher 
particle numbers, as well as higher finite-difference resolution.
If we accept that the metric coefficients become independent of $t$ in 
the critical regime, then the critical spacetime is stationary.
If, in addition, the
vector $N^a=(\partial/\partial t)^a$ is orthogonal to the spatial hypersurfaces, then the spacetime is static, and the 
shift function, $\beta(t,r)$, must vanish.
In Fig. \ref{b} we show the evolution of the shift function.
During the period when the time derivatives of the metric
coefficients are close to zero, the shift function $\beta(t,r)$ is also
close to zero, in the sense that $|\Delta_S ({\beta})| \approx |{\beta}|$.
We thus have evidence that the critical solution in this case is characterized
by a static geometry.

\begin{figure}[htbp]
\epsfxsize=4.0in
\centerline{\epsfbox[100 150 550 700]{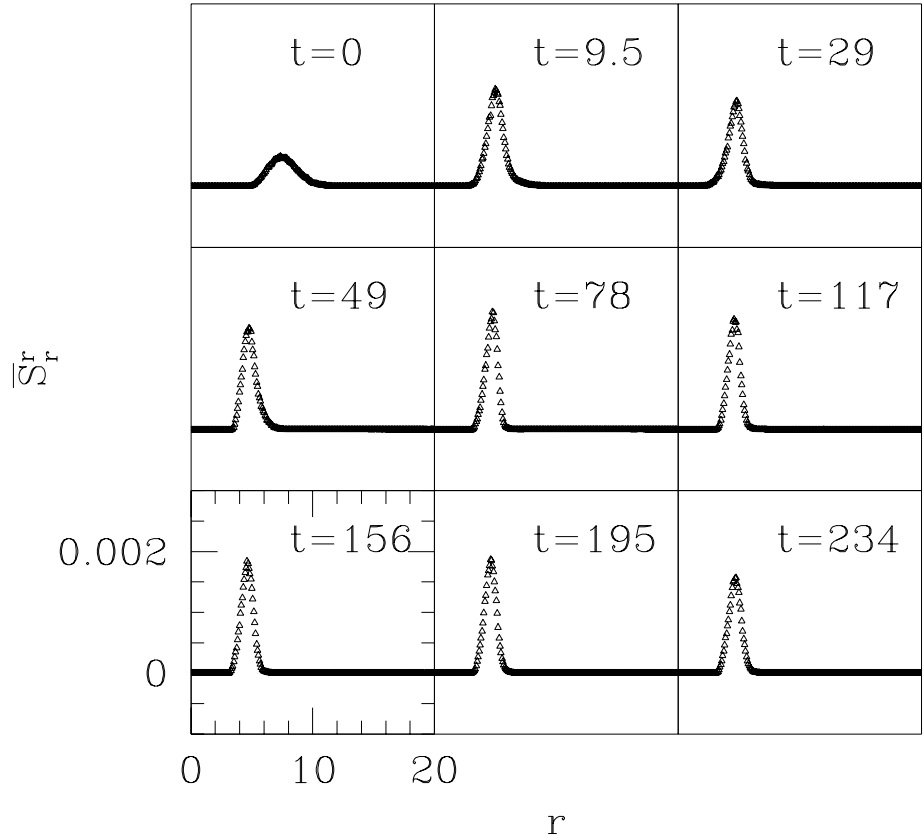}}
\caption[Evolution of $\bar S^r{}_r(t,r)$. During the static regime $\bar S^r{}_r(t,r)$ is not
close to zero showing that although $p_r$ is zero on average $|p_r|$ is not.]
{Evolution of $\bar S^r{}_r(t,r)$ from a marginally subcritical calculation
using Family~(a).
During the static regime, $\bar S^r{}_r(t,r)$ is bounded
away from zero,
showing that although $p_r$ is zero on average, $|p_r|$ is not.}
\label{Sr}
\end{figure}

We have also observed that in the critical regime the total current 
density, $j_r$, tends to zero.  This must 
 be the case if the spacetime is static.  However, as 
shown in Fig.~\ref{Sr}, 
the $\bar S^r{}_r$ component of the stress energy
tensor is {\em non-zero} near criticality.
This means that although on average there are the same
number of particles with positive (outward-directed) and negative 
(inward-directed) $p_r$, the mean value of $|p_r|$ does {\em not} 
vanish.

As is typical of Type I critical solutions, as we tune $M_o \to M_o^\star$,
the dynamical solution spends more and more time ``close'' to the 
putative static solution, and we expect to find power-law scaling of 
the time, $\tau$, (the ``lifetime'' of the near-critical configuration)
spent in the critical regime as a function of $\ln|M_o - M_o^\star|$.
Specifically, we expect the static critical solution (or solutions, since 
we are unable to demonstrate convincingly that the model has a {\em unique}
critical solution, up to trivial rescalings) to possess exactly one 
unstable mode in perturbation theory, with an associated Lyapounov exponent 
which is simply the reciprocal of the scaling exponent, $\sigma$, in the 
lifetime scaling law:
\begin{equation}
\tau \sim -\sigma \, {\mathrm ln}|M_o-M_o{}^\star|
\label{lifetimescaling}
\end{equation} 

\begin{figure}[htbp]
\epsfxsize=4.0in
\centerline{\epsfbox[100 150 550 700]{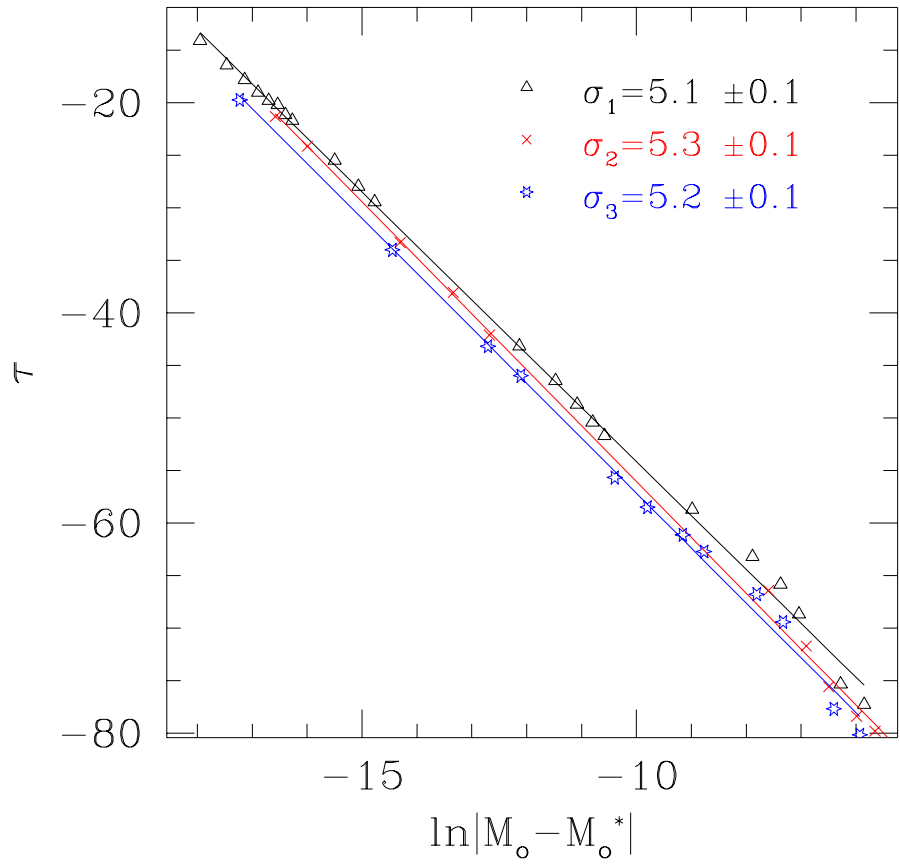}}
\caption[Illustration of scaling law for the lifetime of near-critical
configurations. The quoted uncertainty for each value of $\sigma$
is the standard deviation of the slope which has been computed
using a least squares fit.]{Illustration of scaling law for the lifetime of near-critical
configurations. The quoted uncertainty for each value of $\sigma$
is the standard deviation of the slope, which has been computed
using a least squares fit.
}
\label{scaling}
\end{figure}

Fig. \ref{scaling} shows a
plot of $\tau=t-t_c$ versus ln$|M_o-M_o^\star|$ 
where $t$ is the total time that the particles in the solution 
generated with parameter $M_o$ are
localized within
$r=r_o=6$, and $t_c$ is the same quantity for
the solution closest to criticality.
We show results from calculations using three different sets of particles 
and the same gaussian family previously discussed.
Using the residual scaling freedom in the model (the equations of motion 
are invariant under $t\to \kappa t$, $r \to \kappa r$ for arbitrary 
$\kappa > 0$), we have 
also normalized each critical solution 
to have unit ADM mass:
\begin{eqnarray}
\label{sca1}
r &\rightarrow& r /M^c(t^\star,r_{\mathrm max})\\
\label{sca2}
t &\rightarrow& t /M^c(t^\star,r_{\mathrm max}),
\end{eqnarray}
Here $t^\star$ is defined to be the instant at which the time 
derivatives of the metric components are closest to zero for the 
solution closest to criticality.
$M^c(t^\star,r_{\mathrm max})$ is then the value of the mass aspect function
at $t=t^\star$, $r=r_{\mathrm max}$, again for the most nearly-critical solution.
We can see that there is a {\it rough} linear relation between the
lifetime $\tau$ of the near-critical configurations and 
$\ln|M_o-M_o^\star|$:
\begin{equation}
\label{scal_eqn}
\tau \sim  -(5.2 \pm 0.2) \ln|M_o-M_o{}^\star|
\end{equation}
where the quoted uncertainty is an estimate of the statistical error.

Qualitatively, then, our results are similar to what has been observed in 
other instances of Type I critical collapse~\cite{TypeI}. 
We have also found results similar to those just presented by using
4 other gaussian families (Table~\ref{table}, Families (b)--(e)), each 
with $\bar p_r{}_o=-4$ and with varying $l_o$'s of 3, 5, 7
and 12; $r_o=5$, $\Delta_{r}=1$, $\Delta_{\bar p_r}=2$ 
and $\Delta_{l}=2$, as for Family~(a)). 
Specifically, in each case we find that the critical geometry appears 
to be static. 
We note that for smaller values of $l_o$, the mass in the critical solution
gets increasingly concentrated near $r=0$, making accurate evolution 
with a uniform finite-difference grid more difficult.
Finally, we have studied a family with the following initial single-particle
distributions (Table~\ref{table}, Family~(f)):
\begin{eqnarray}
\label{tanh1}
R(r)&\propto&\left(1-\tanh\left((r-r_o)/\Delta_r\right)^2 \right) \, \Theta (r) \\
\label{tanh2}
P(p_r)&\propto&\left(1-\tanh\left((\bar p_r-{\bar p_r}{}_o)/\Delta_{\bar p_r}\right)\right)\\
\label{tanh3}
L(l)&\propto&\left(1-\tanh\left((l-l_o)/\Delta_l\right)^2 \right) \,\Theta (l).
\end{eqnarray}
Here we took $r_o=5$, $\Delta_r=1$, ${\bar p_r}{}_o=-4$, 
$\Delta_{\bar pr}=2$, $l_o=7$ and $\Delta_l=2$.
For this data we also find evidence that as 
$M_o \to M_o^\star$, the geometry becomes static.

\begin{figure}[htbp]
\epsfxsize=4.0in
\centerline{\epsfbox[100 150 550 700]{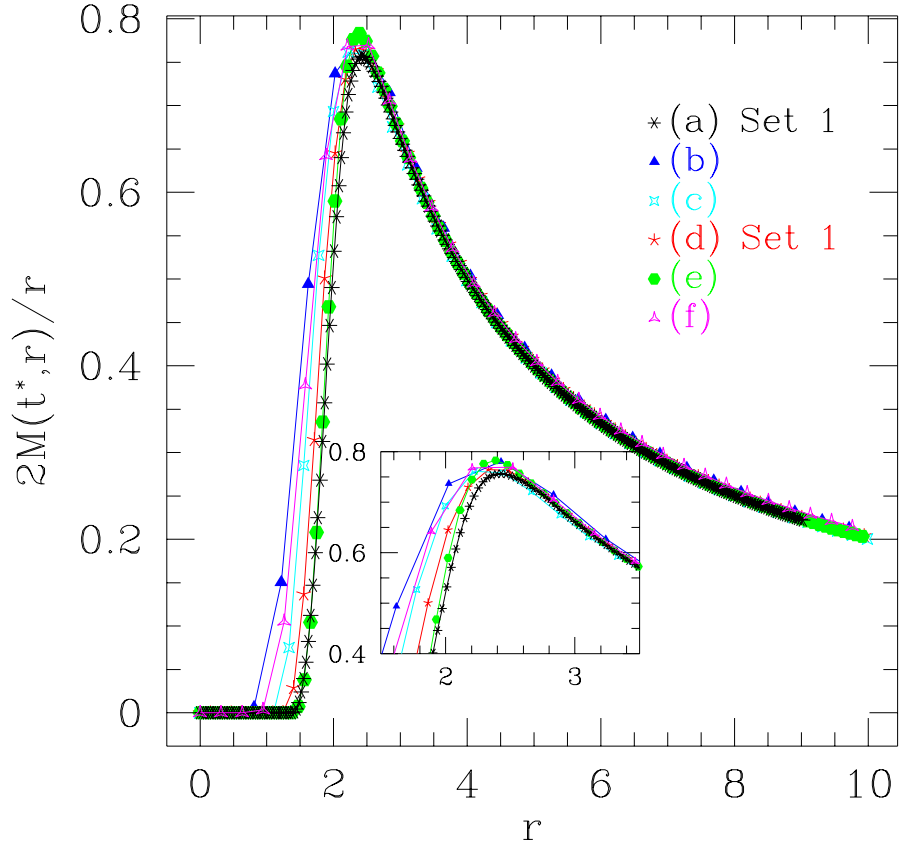}}
\caption[Critical solution for different families of initial data. We can see that there is
agreement on the profiles of $2M(t^\star,r)/r$ where $t^\star$ is the
time when the solution is in the near-critical evolution.]
{Comparison of near-critical solutions for different families of
initial data (Table~\ref{table}).
We see
evidence for a universal profile  $2M(t^\star,r)/r$,
where $t^\star$ (different for each family) is the instant when
the temporal derivatives of the metric components are minimized,
and $r$ has been rescaled for each family so that all critical 
solutions have unit ADM mass.
The maximum value of $2M(r)/r$ is about $0.76$
showing immediately that the critical solution cannot be one of the
clusters considered in \cite{Einstein:1939}, since there are no
equilibrium Einstein clusters with maximum $2M(r)/r$ larger than $2/3$.
(Moreover, in contrast to the configurations studied here, all particles
in an Einstein cluster are in circular orbits.)}
\label{families}
\end{figure}

In Fig.~\ref{families} we show profiles of  $2M(t^\star,r)/r$ for 
all of the different
families considered, each separate profile being selected from the 
corresponding period of near-critical evolution. Again, since different 
initial conditions set different overall length scales for the problem, 
we have normalized the results using the rescaling given by
equations~(\ref{sca1}-\ref{sca2}).  We see that, after normalization, 
the peak of $2M(t^\star,r)/r$ is {\em roughly}
at the same radial location, $r \equiv r^\star=2.3$. We also find that 
the better resolved a solution is, the closer it conforms to the best 
resolved solution (Table~\ref{table}, Family~(a)).
This provides some indication that there may
be a {\em universal} critical solution in this model (up to trivial 
rescalings, $r \to \kappa r$, $t \to \kappa t$), but again, we 
would need better finite-difference resolution and many more particles to 
verify this conjecture. This figure also shows that the maximum of $2M(t^\star,r)/r$
has a value of approximately $0.76$. This immediately shows that the critical solution
is not one of the clusters considered in
\cite{Einstein:1939}, since there are no equilibrium Einstein clusters with maximum $2M(r)/r$ 
larger than $2/3$.

\begin{figure}[htbp]
\epsfxsize=6.0in
\centerline{\epsfbox[35 300 520 647]{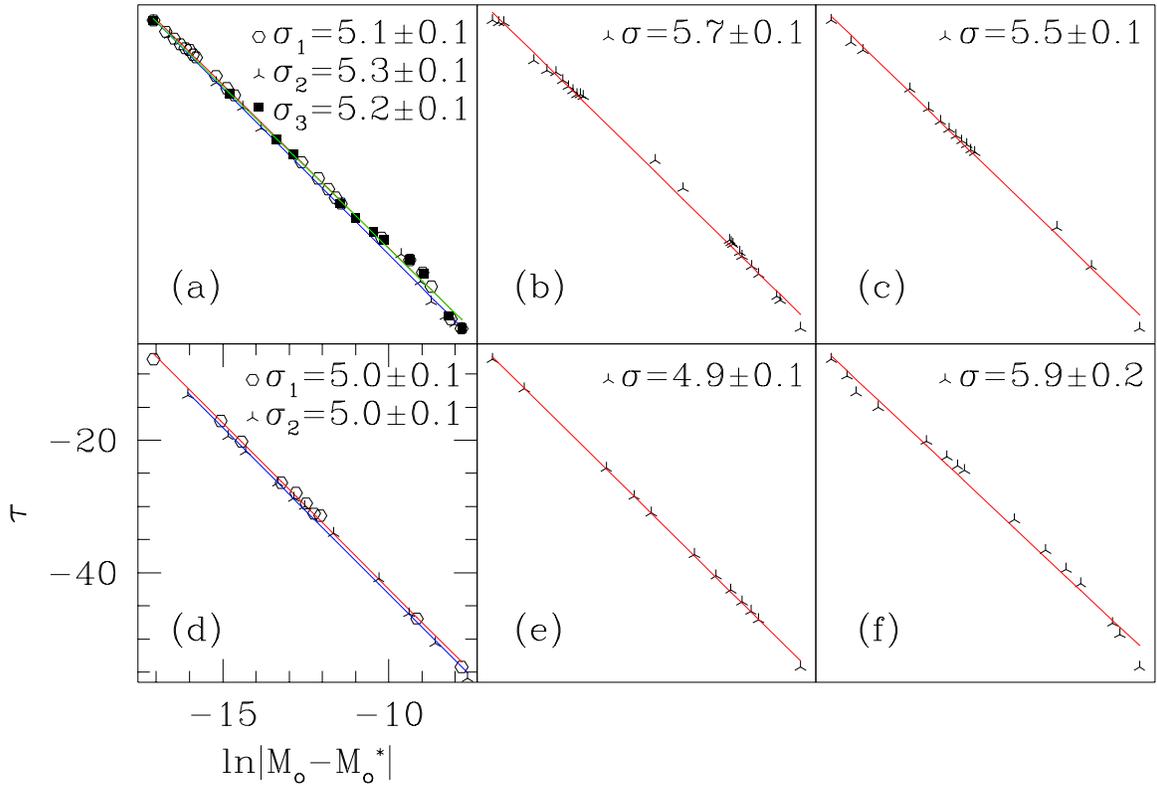}}
\caption[Scaling behavior for different families of initial data. We observe
near-critical lifetime-scaling
behavior for all the families we have studied, as expected for Type I
solutions.
The uncertainty on $sigma$ is given by the
stander deviation of the slope computed from a least squares fit.]
{Scaling behavior for different families of initial data.
We observe
near-critical lifetime-scaling
behavior for all the families we have studied, as expected for Type I
solutions (static or periodic solutions, with {\em one} unstable mode
in perturbation theory).
The quoted uncertainty in $\sigma$ is given by the
standard deviation of the least-squares slope.  The axes ranges vary 
somewhat from sub-plot to sub-plot; the values shown for Family~(d)
are representative.
}
\label{sigmasplot}
\end{figure}

We have also estimated $\sigma$ defined by equation~(\ref{lifetimescaling})
for the different families described above.
Fig. \ref{sigmasplot} shows the lifetime scaling measured for the
various initial data sets, where the quoted uncertainty in 
each value of $\sigma$ is the standard deviation of the slope 
computed from a least squares fit. The values that 
we have obtained for the scaling exponents are 
also collected in Table~\ref{table}.

\begin{figure}[htbp]
\epsfxsize=4.0in
\centerline{\epsfbox[100 150 550 700]{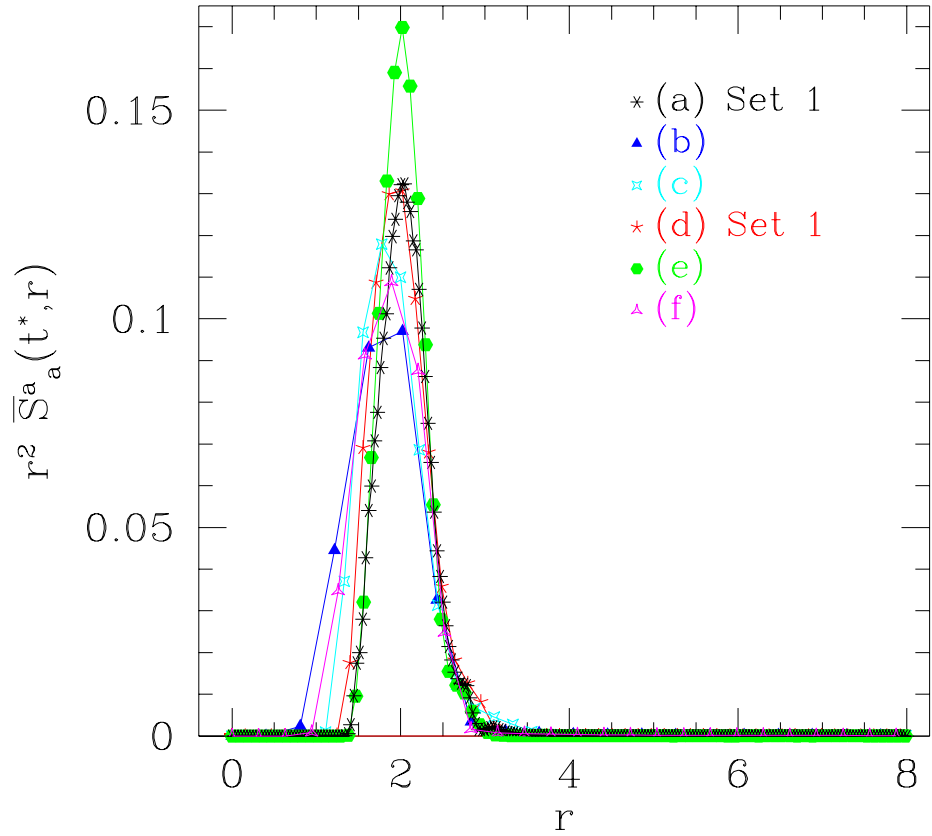}}
\caption[$r^2 \bar S^a{}_a(t^\star,r)$ for the different families during the critical regime.]
{$r^2 \bar S^a{}_a(t^\star,r)$ for the different families in the near-critical regime.
As described in the text, this quantity is a dimensionless measure of
the squared-angular-momentum of the distribution.  In contrast to
Fig.~\ref{families}, we see no particular
evidence of a universal profile here.  However, considerably more
resolution (both in $h$ {\em and} $N$) is needed to 
accurately assess the impact of angular momentum on critical collapse 
in this model.
}
\label{r2sa}
\end{figure}

Finally, we are also interested in investigating the dependence of 
the critical solutions on the 
distribution of angular momentum.
In Fig. \ref{r2sa} we show $r^2\, \bar S^a{}_a(t^\star,r)$ for the different families of 
initial data we have studied.  Here
\begin{equation}
	r^2\, \bar S^a{}_a = r^2\left[\sum_{i=1}^N \, W(r - r_i) \,
                        \frac{1}{4\pi\Delta r} \frac{\left[l\right]^2_i}{r^4_i \left[\bar p^t\right]_i} \right]\, 
\end{equation}
(see equation (\ref{sr_sa})), and $t^\star$ is defined as previously.
We note that
$r^2 S^a{}_a(t,r)$, is a dimensionless quantity which measures the square
of the angular momentum
of the
distribution of particles. As in~Fig. \ref{families}, we have again rescaled the radial 
coordinate (and time) so that the critical configuration has unit ADM mass.
We see that there is no obvious agreement of the
profiles calculated from different families of initial data; clearly 
more work needs to be done in order to clarify the effect of initial
angular momentum distributions on critical evolution in this model.

\section{Conclusions}
\label{sec:discussion}

We have studied critical behavior at the threshold of black hole formation for 
collisionless matter with angular momentum and have corroborated 
the findings of Rein~{\em et al}~\cite{Rein} that the black holes 
which form at threshold 
in this model are of {\em finite} mass (Type I behavior).
Further, our results indicate that for families with non-zero angular 
momentum, 
the critical solution has a static geometry,
with non-zero radial particle momenta. 
We have also found evidence for a lifetime scaling law which is to 
be expected for Type I critical solutions, and have some indications 
of universality.
In order to produce more definitive results using our current 
approach, we would need to employ {\em many} more particles and 
better finite-difference resolution.   Since the critical behavior 
in this model does {\em not} appear to generate structure on 
arbitrarily small scales, it seems unlikely that adaptive methods,
such as those used in~\cite{Choptuik}, would be of much help here.
Thus,
it may be that the development of a finite-difference code to solve the 
Vlasov equation directly in phase space
would be the best route to more accurate results. 
Perhaps most importantly, 
this should provide a technique with better-understood, and 
better-controllable, convergence properties.

\section*{Acknowledgments}
\label{sec:ack}
It is our pleasure to thank W.G.~Unruh, L.~Lehner, and the rest 
of the members of the numerical relativity group at
the University of British Columbia for many useful discussions.
This research was supported by NSERC, the Canadian Institute for 
Advanced Research and the 
Government of the Basque Country through a fellowship to I.O.
Most of the calculations were performed on the  
{\tt vn.physics.ubc.ca} Beowulf cluster, which was funded by the 
Canadian Foundation for Innovation.

\vbox{
\begin{table}[htbp]
\begin{center}
\begin{tabular}{clcc}\hline
Family & Form of Initial Data  & Set No. & $\sigma$ \\ \hline
 (a) & Gaussian,\quad $l_o=12$ ATS & 1 & $5.1\pm0.2$ \\
 (a) & Gaussian,\quad $l_o=12$ ATS & 2 & $5.3\pm0.2$ \\
 (a) & Gaussian,\quad $l_o=12$ ATS & 3 & $5.2\pm0.2$ \\ \hline
 (b) & Gaussian,\quad $l_o=3$  & 1 & $5.7\pm0.2$ \\ \hline
 (c) & Gaussian,\quad $l_o=5$  & 1 & $5.5\pm0.2$ \\ \hline
 (d) & Gaussian,\quad $l_o=7$  & 1 & $5.0\pm0.2$ \\
 (d) & Gaussian,\quad $l_o=7$  & 2 & $5.0\pm0.2$ \\ \hline
 (e) & Gaussian,\quad $l_o=12$ & 1 & $4.9\pm0.2$ \\ \hline
 (f) & Tanh,\quad $l_o=7$ &  1 &$5.9\pm0.2$ \\
\end{tabular}
\end{center}
\caption[~]{
Summary of critical searches described in the text.  Listed are family label, form
of initial data, set number (for families where multiple, independent, $N$-particle representations of
the initial distribution function were used) and computed lifetime-scaling exponents, $\sigma$.
See the text and particularly equations~(\ref{gaussian1}-\ref{gaussian3}) and
(\ref{tanh1}-\ref{tanh3}), respectively, for detailed definitions of the ``Gaussian'' and ``Tanh'' initial
data.  Also note that ATS stands for ``almost time symmetric'', as discussed
in the text.  The quoted error for {\em all} values of $\sigma$ is the estimated statistical error for
the ATS data (see equation (\ref{scal_eqn})).
}
\label{table}
\end{table}
} 
\begin{figure}[htbp]
\epsfxsize=4.0in
\centerline{\epsfbox[0 0 535 410]{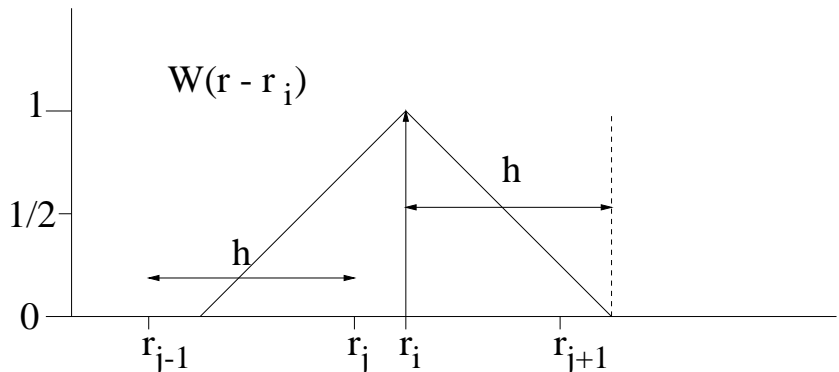}}
\caption[$W(r-r_p)$.]
{Illustration of the smoothing kernel, $W(r-r_i)$. Here, $r_i$ is the
position of a particle; $\cdots\,r_{j-1},\,r_j,\,r_{j+1}\,\cdots$ are
the (uniform) finite-difference mesh points with $\Delta_r = r_{j+1} - r_j
= {\rm constant} = h$}
\label{cic}
\end{figure}
\end{document}